# STUDY OF HIGH PRESSURE GAS FILLED RF CAVITIES FOR MUON COLLIDER

Katsuya Yonehara

*Fermi National Accelerator Laboratory, Batavia, IL 60510, USA*

**Abstract.** Muon collider is a considerable candidate of the next-generation high-energy lepton collider machine. Operating an RF cavity in a multi-Tesla magnet is a critical requirement in a muon accelerator and a cooling channel. However, the maximum RF gradient in a vacuum RF cavity is strongly limited by an external magnetic field. Dense hydrogen gas filled RF cavity has been proposed since it is functional of generating a high RF accelerating gradient in a strong magnetic field and making an ionization cooling process at the same time. A critical issue of the cavity is a beam-induced plasma that consumes a considerable amount of RF power. The gas filled RF test cell was made and measured the RF loading due to a beam-induced plasma by using an intense proton beam at Fermilab. By doping an electronegative gas in dense hydrogen, the plasma loading effect is significantly mitigated. The result shows that the cavity is functional with a muon collider beam. Recent progress is shown in this presentation.

## INTRODUCTION

### Why Muon Collider?

A next-generation high-energy lepton collider project needs to be developed to follow up LHC (Large Hadron Collider). To this end, an electron-positron collider, e.g. ILC (International Linear Collider) and CLIC (Compact LInear Collider), has been investigated. The size (length) of an electron-positron collider is enormous even though they could use a state of the art technology of high gradient RF accelerator because of a synchrotron radiation loss. On contrary, muons do not have such an intrinsic issue because they are 200 times heavier than electron. Therefore, muons can be accelerated up to 0.75~1.5 TeV energy in a circular accelerating machine. In fact, the size of 1.5 TeV CoM muon collider will be small enough to fit in the present Fermilab site [1]. Table 1 shows the designed muon beam parameter in the collider ring.

**TABLE 1.** Beam parameter in the muon collider ring [2].

| Parameter | | |
|---|---|---|
| $\sqrt{s}$ (TeV) | 1.5 | 3.0 |
| $\beta^\star$ (cm) (bare lattice) | 1 (0.5-2) | 0.5 (0.3-3) |
| Av. Luminosity/IP ($10^{34}$ cm$^{-2}$ s$^{-1}$) | 1.25 | 4.4 |
| Max. bending field (T) | 10 | 10 |
| Av. bending field in arcs (T) | 8.3 | 8.4 |
| Circumference (km) | 2.5 (2.7) | 4.45 |
| No. of IPs | 2 | 2 |
| Repetition rate (Hz) | 15 | 12 |
| Beam-beam parameter/IP | 0.087 | 0.087 |
| Beam size /IP (μm) | 6 | 3 |
| Bunch length (cm) | 1 | 0.5 |
| No. muons/bunch ($10^{12}$) | 2 | 2 |
| Energy spread (%) | 0.1 | 0.1 |

### Required RF Performance in Muon Cooling

Since muons are tertiary particles, a volume of muon beam phase space after a pion decay channel is too large to be accepted in a conventional RF accelerator. It requires muon beam phase space cooling. Besides, muons have a finite lifetime that is 2.2γ μs. Fortunately, muon lifetime is long enough to condition the muon phase space by using a novel cooling technique, which is called ionization cooling. The ionization cooling theory is similar as the electron cooling one. A heat transfer takes place between a projectile (muon beam) and a cooling object (ionization cooling material) via a Coulomb interaction. A great advantage of ionization cooling is that the process is several orders of magnitude quicker than the electron cooling since more electrons are provided in the ionization cooling material. A cooling decrement in an ideal case is estimated,

$$\varepsilon_n = \varepsilon_0 \exp\left(-\frac{\langle dE/dx \rangle}{\beta^2 E} z\right) \quad (1)$$

where $\beta$ is a relativistic particle velocity and $E$ is a kinetic energy of beam. $\langle dE/dx \rangle$ is an ionization energy loss rate. If an RF accelerator fully recovers the ionization energy loss, $\langle dE/dx \rangle$ can also represent the RF acceleration gradient. Figure 1 shows the required length of ionization cooling channel as a function of the RF recovery acceleration field gradient. The goal phase space cooling factor for a high energy muon collider is $10^6$. If the available RF recovery acceleration field gradient is 10~15 MV/m, the length of cooling channel can be 150~250 m.

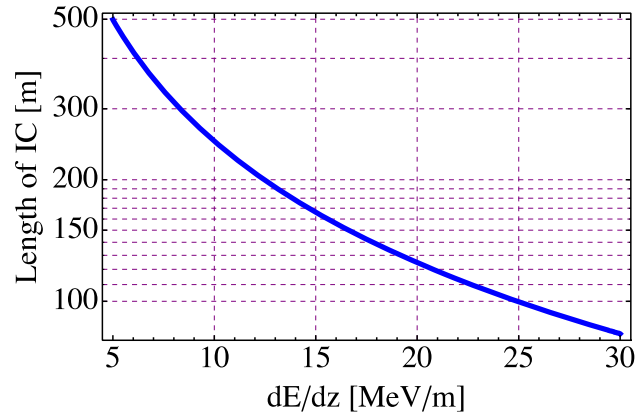

**FIGURE 1.** Required length of ionization cooling channel as a function of RF regain field gradient to achieve a design cooling factor $10^6$.

There is another scattering process between a projectile and nuclei of an ionization cooling material, which makes a large scattering angle (Rutherford scattering), which is called a multiple scattering process. A transverse beam phase space is inflated due to the multiple scattering process. As a result, the emittance evolution of transverse beam phase space in a cooling channel is determined by a balancing between a cooling and heating processes. Thus, the normalized equilibrium beam emittance in a transverse phase space is given from the balancing equation,

$$\bar{\varepsilon}_{normal} = \frac{\beta_t (13.6 MeV)^2}{2\beta m_\mu L_R} \left\langle \frac{dE}{dx} \right\rangle^{-1} \quad (2)$$

where $L_R$ is a radiation length of cooling material and $\beta_t$ is a transverse beta function. This relation suggests that the best ionization cooling material must have a large ionization energy loss and a long radiation length (i.e. a low Z material) and a beam cooing lattice must produce a low beta function at a cooling material. Liquid hydrogen is usually selected as the ionization cooling material for this reason.

## RF Cavity in Ionization Cooling Channel

Figure 2 shows a schematic layout of an ionization cooling channel. To make a small $\beta_t$, a solenoid magnet is preferred to be used because it can focus a beam simultaneously in both x and y planes. An ionization cooling absorber locates at the lowest beta function to maximize the momentum transfer efficiency via the ionization loss process. In order to recover the ionization energy loss, an RF cavity is adjacently located either in front or behind the ionization cooling material. It is worth to emphasize that the RF cavity should be embedded in a strong magnetic field to have a compact channel. The field strength at the RF cavity is 3 Tesla and higher.

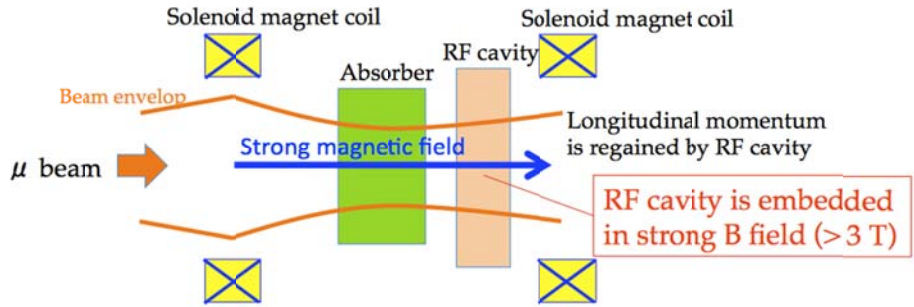

**FIGURE 2.** Schematic drawing of ionization cooling channel.

Available maximum RF gradient in a vacuum RF pillbox cavity is limited by the strength of magnetic field. The magnetic field concentrates a dark current density in the cavity [3,4]. Consequently, the probability of RF breakdown that is ignited by the dark current is higher in stronger magnet. In past 805 MHz vacuum pillbox RF cavity test, the available maximum RF field gradient was degraded by 1/3 at 3 Tesla from the field gradient at zero magnetic field. This value is about a half smaller than the desired RF field gradient for a cooling channel.

## Gas Filled RF Cavity

By filling a dense buffer gas in a RF cavity, the dark current flow can be diffused by a Coulomb scattering. Molecular hydrogen is the ideal buffer gas since it produces the lowest multiple scattering angle in the ionization loss process. Besides, a dense hydrogen gas can be used as the best ionization cooling material as discussed in previous section. A high pressure gas filled RF test cell was demonstrated in a strong magnetic field and found no RF field degradation by the magnetic field [5]. However, a dense beam-induced plasma is generated in the gas filled RF cavity that makes a huge RF power loading effect. It is called the beam-plasma loading effect. The beam-induced plasma is shaken by the RF field, and takes energy from the field and loses it by collision of its ions and electrons with the gas molecule. Consequently, the cavity $Q$ factor is degraded as a function of the beam-induced plasma density. According to the model, ionized electrons are a dominant part of plasma loading effect because of its light mass. It suggests that the RF loading effect could be significantly reduced by doping an electronegative gas.

There is no intense muon source to test a gas filled RF cavity with a muon collider beam. Fortunately, the beam-induced plasma model can be experimentally validated by using a standard charged beam. If the model is well-established, we can extrapolate the loading parameter to the muon collider beam. A dense gas filled RF cavity beam test has been carried by using a 400 MeV proton beam in the MTA (Mucool Test Area) at Fermilab in Summer 2011 and Spring 2012. We mainly report the former experimental result and analysis in this document.

## EXPERIMENT

Details of the experiment are depicted in Refs. 6,7 and 8. Here, we describe the experimental parameters that are needed for data analysis. A 400 MeV $H^-$ beam was delivered from an extended Fermilab Linac beam line to a high pressure gas filled 800 MHz RF test cell. An $H^-$ is fully stripped at a vacuum and pressure windows and turned to a proton. The beam intensity in the cavity was tuned by changing the beam spot size at the input to a collimator with a 4 mm diameter hole located upstream of the cavity. Beam current in the cavity was measured by using two toroid beam current monitors. The total number of beam bunches was 1500 with a 5 ns bunch spacing and a total length 7.5 μs. These beam elements were located in a 3-Tesla solenoid magnet. The shunt impedance of the cavity ($R$) was 2.0 MΩ at 1 atm. The cavity was powered by a 13 MW Klystron. Since a lifetime of beam-plasma loading effect is longer than a 200 MHz beam bunch spacing the phase of 800 MHz cavity was not locked to that of the beam. There was an RF pickup loop in the cavity to monitor RF field strength.

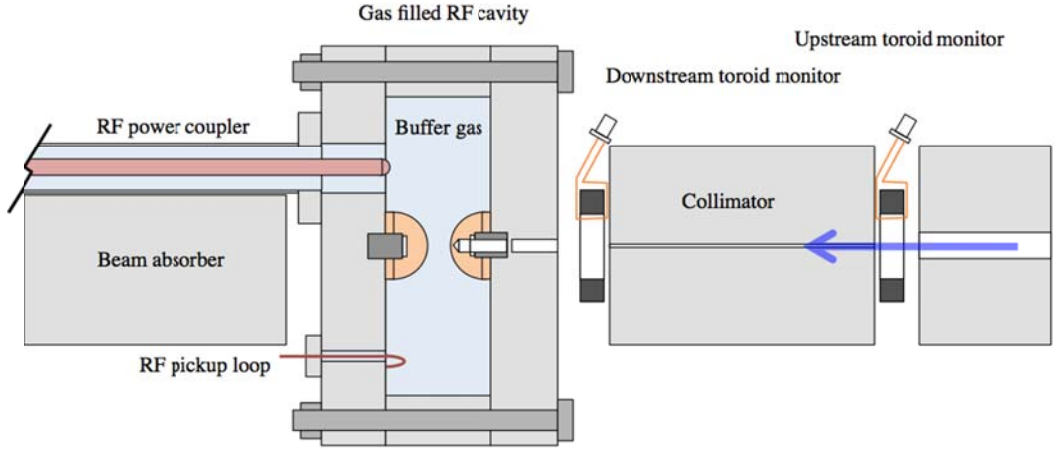

**FIGURE 3.** Observed RF pickup signal (red) and toroid current monitor signal (blue).

First, RF energy consumptions of single electron (*dw*) in pure hydrogen and nitrogen gases were investigated with various peak RF field gradients (*E*) and gas pressures (*p*). The observed *dw* is well agreed with the beam-plasma loading effect model. It suggests that the model can be valid to estimate the RF Q factor reduction as functions of $E/p$, beam intensity, and time structure. Figure 4 shows a typical RF pickup signal and a toroid signal. It should note that the test cell did not breakdown under an intense beam although the beam-plasma loading effect was dominant, i.e. a rapid RF amplitude drop when the beam was on. After the beam was turned off, the RF power was restored.

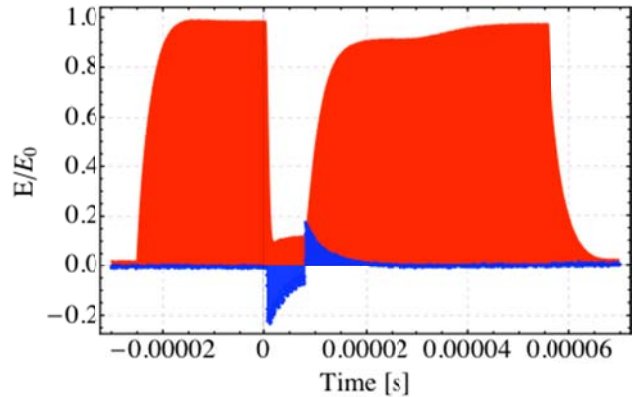

**FIGURE 4.** Observed RF pickup signal (red) and toroid current monitor signal (blue).

## Analyze Data

*Solve equivalent resonant electric circuit model*

An RF system can be represented as an *LCR* resonant circuit diagram as shown in Figure 5. Thevenin's Theorem has been used to combine the cavity load and the generator resistance, i.e. $V_0 = V_g/2$ and if the cavity is matched $Z = R$ without beam loading. When a beam appears in the cavity, a new resistance, $R_p$ is added in the equivalent circuit. It represents the conductance of beam-induced hydrogen plasma in the cavity.

The power balance equation is given by:

$$p_{plasma} = \frac{(V_0 - V)V}{R} - \frac{d}{dt}\left(\frac{CV^2}{2}\right) \qquad (3)$$

where $p_{plasma}$ is the RF dissipated power in a plasma and $V_0$ is the peak initial field potential in the cavity without beam and *V* is the measured RF amplitude. The first term in the right hand side equation is the RF power gain from the Klystron and the second one power lost or gained from the cavity when its peak voltage changes.

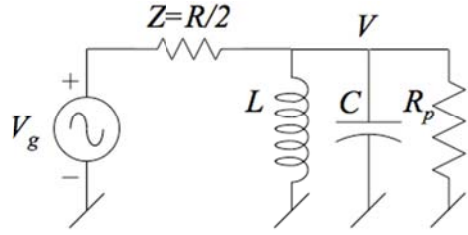

**FIGURE 5.** RF resonance circuit diagram with beam-plasma loading.

The ionization energy loss of a proton in a matter is well reproduced by the Bethe-Bloch formula. Since the ion-electron pair production energy ($W$) in hydrogen and nitrogen are known 35.3 and 33.6 eV, respectively [9], the number of electrons in the cavity produced by single proton can be estimated by

$$n_e = \left\langle \frac{dE}{dx} \right\rangle \frac{\rho}{W} \qquad (4)$$

where $\langle dE/dx \rangle$ is the Bethe-Bloch formula, $\rho$ is the density of gaseous hydrogen.

The time-domain rate equation of the number of electrons in a pure hydrogen gas is

$$\frac{dn_e}{dt} = \dot{N} - \beta n_e n_{ion} \qquad (5)$$

where $\dot{N}$ is a production rate of electrons due to the ionization process, $\beta$ is the hydrogen recombination rate. The electron diffusion rate is small and is omitted from Eq. (3). Because the hydrogen recombination process goes as $n^2$ it is zero at the beginning of beam-on, hence the $dn_e/dt$ is simply represented as the electron production rate $\dot{N}$. Then, the RF dissipation energy of single electron, $dw$ can be estimated from Eqs. (1) and (2),

$$dw = \frac{P_{plasma}}{dn_e/dt} \qquad (6)$$

The $dw$ in $H_2$ was taken with wider variables range. Figure 6 shows the $dw$ as a function $E$ in $p = 500, 800, 950$ psi. The $dw$ is extracted in $N_2$ and $H_2$ gases, which are $2 \cdot 10^{-17}$ and $3 \cdot 10^{-17}$ Joules/RF cycle at $E = 20$ MV/m, respectively. There is a strong correlation between $dw$ and $E$. A line in each plot shows the best chi-square fit of the data by using an equation $aE^b$. We found that the fitting also has a small dependence upon the gas pressure. On the other hand, there is no beam intensity dependence on the $dw$. In fact, a point at the same $E$ in the plot is taken with different beam intensity.

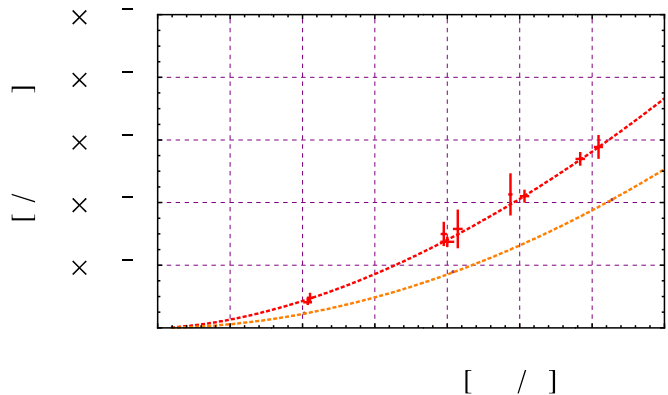

**FIGURE 6.** $dw$ in $N_2$ and $H_2$ at 500 psi as a function of $E$. Error bars show the statistic error.

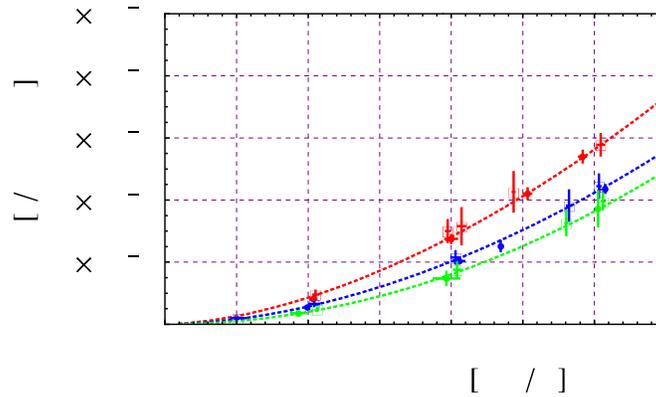

**FIGURE 7.** Observed *dw* as a function of *E*. Red, blue, and green marks are *dw* in 500, 800, 950 psi, respectively. Error bars show the statistic error.

*Estimate RF energy dissipation per single electron*

The RF energy dissipation per single electron, *dw* can be estimated by using a simple electron transport analysis. Since the electron drift velocity in $H_2$ gas is known [10], the total RF power dissipation can be given by

$$p_{model} = jE = qn_e u_e E \tag{7}$$

where *j* and *E* are a current and an electric field gradient, respectively, and $u_e$ is the drift velocity of electron in a matter. $u_e$ is a function of *E/p*. Therefore, in order to estimate *dw*, eq. (7) must be integrated as a function of time,

$$dw = 2\int_{T/2} qn_e u_e \left(E_0 \sin(2\pi vt), p\right) E_0 \sin(2\pi vt) dt \tag{8}$$

where $E_0$ is a peak RF gradient, *v* is an RF resonant frequency, *T* is an RF period, and *p* is a gas pressure. The observed *dw* in this experiment is smaller than the estimated *dw* that is given from eq. (8). Figure 8 shows the correction factor of *dw*. The correction factor becomes larger in higher gas pressure. It indicates that the correction factor has gas density dependence.

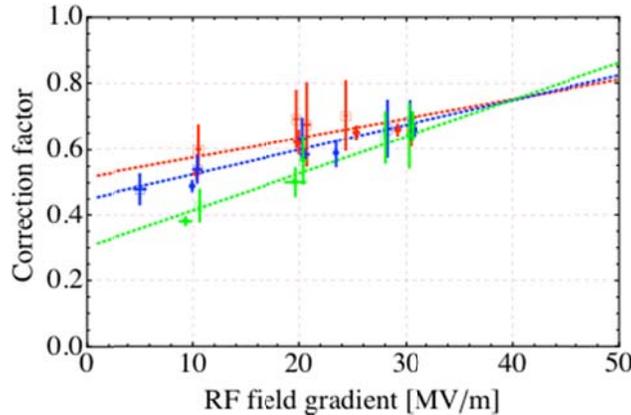

**FIGURE 8.** Correction factor of *dw*. Red, blue, and green marks are 500, 800, and 950 psi pure $H_2$ gases, respectively. Error bars show the statistic error.

*Electronegative gas effect*

Electronegative gas effect was studied. Since the electron capture cross section of $SF_6$ is well-known, $SF_6$ is uses as a reference electronegative gas. However, $SF_6$ is a nasty material. $SF_6$ is disintegrated and turned into Fluorine in a breakdown process. Molecular oxygen is a good candidate for a practical electronegative gas for an ionization cooling channel since it has a large electron capture cross section via the three body reaction ($A + B + e^- \rightarrow A^- + B$). But, the flammable gas hazard should be concerned in which the concentration of $O_2$ should not exceed the LFL

(Lowest Flammable Level). Therefore, a DA (Dry Air) was used in which $O_2$ abundance is 20 %. From past experiment, we knew that $N_2$ doped $H_2$ gas does not remove the ionized electron from the cavity even $NH_3$ would be formed from ½ $N_2$ + 3/2 $H_2$ reaction.

The RF amplitude drop has been observed in various gas combinations, i.e. SF6, He and $N_2$. Figure 9 shows the initial RF amplitude drop in various buffer gases. The RF amplitude drops in 1 % DA doped $H_2$ and 0.01 % SF6 doped $H_2$ are very similar while that in 1 % DA doped He gas is steeper and that in 1 % DA doped $N_2$ gas is shallower than the $H_2$ based dopant gas. From these evidences, we brought up the model that the residual ions, i.e. positive hydrogen and negative ions, must contribute the RF power dissipation from the cavity since there is no electron in the cavity.

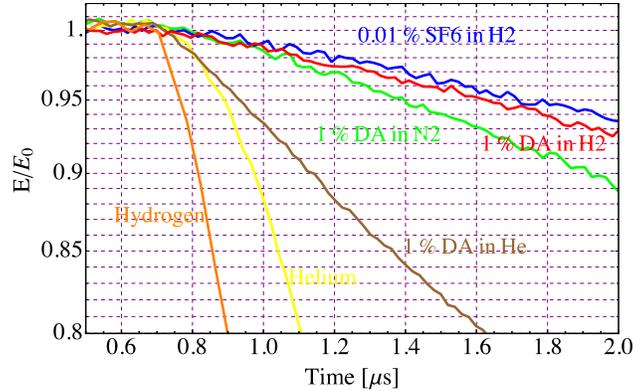

**FIGURE 4.** Initial RF amplitude drops in various gas condition. (Blue) 0.01 % $SF_6$ in $H_2$ gas, (red) 1 % DA in $H_2$ gas, (green) 1 % DA in $N_2$ gas, (brown) 1 % DA in He gas, (yellow) pure He gas, and (orange) pure $H_2$ gas, respectively. Gas pressure was 1470 psi in all cases except for $N_2$ gas.

Figure 10 shows the summary of observed *dw* in various gases as a function of E/p. We can clearly see the tendency of *dw* at different gases. As we expected, the highest RF power dissipation takes place in He parent gas and the lowest one is in $N_2$ one. An orange line in Figure 11 is the estimated *dw* if it is predominated by $H_5^+$ (reduced μ = 9.6 cm$^2$/V/s). The experimental data seems to be reproduced by the model. It should note that the discrepancy between experiment and model is larger in higher gas pressure. We do not fully understand the physics yet.

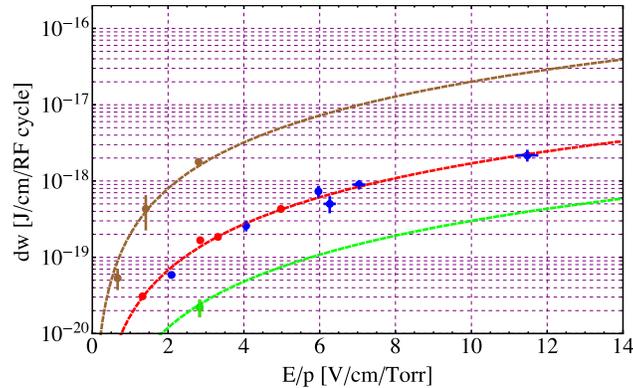

**FIGURE 5.** Observed dw in various gases. (Brown) 1 % DA in 1470 psi He, (red) 1 % DA in 1470 psi H2, (blue) 0.01 % SF6 in 500, 800, and 950 psi H2, and (green) 700 psi N2 gases, respectively.

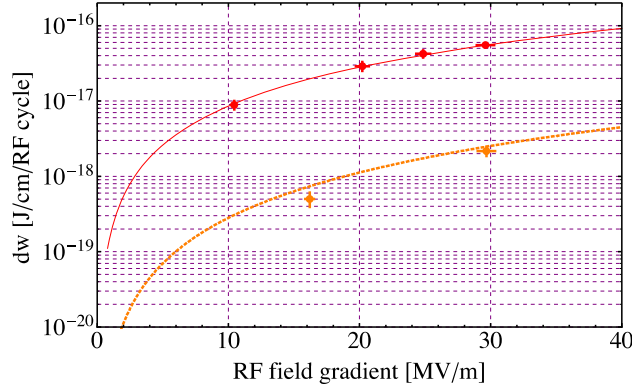

**FIGURE 6.** Compare *dw* in pure $H_2$ (red) and *dw* in 0.01 % $SF_6$ doped $H_2$ (orange) at 500 psi.

# ESTIMATE BEAM-PLASMA LOADING EFFECT WITH MUON COLLIDER BEAM PARAMETER

The beam-plasma loading effect in a realistic high pressure gas filled RF pillbox cavity is estimated. A 4 MW 8 GeV proton beam is generated in a proton driver [11]. In order to maximize pion/muon capture rate in a muon beam frontend channel, proton beam is accumulated and compressed in 2 ± 1 ns in a bunch compressor ring. The protons would be formed into 2ns-long bunches that hit the target at 15 Hz (2.1 $10^{14}$ /pulse). The pion's from that collision would be captured by the front end transport and RF into a series of $\mu^+$ and $\mu^-$ bunches that will propagate through the cooling channel. 12 $\mu^-$ bunches are obtained in 200 MHz spacing of varying intensity (with fewer muons toward later bunches), and one will also have a similar train of $\mu^+$ bunches. For a first estimate of the resulting secondary beam, we estimate that each proton would produce ~0.2 $\mu^\pm$ and that these are split into 12 bunches spaced by 5ns; in this model there would then be 3.5 $10^{12}$ $\mu^\pm$ charges per bunch. Therefore, 4.2 $10^{13}$ μs go through the cavity in 60 ns.

The stored energy of RF pillbox cavity can be calculated from

$$U = \frac{\varepsilon}{2} E_0^2 \left( J_1(2.405) \right)^2 \pi R_c^2 \qquad (9)$$

where $\varepsilon$ is a dielectric constant, $E_0$ is a peak RF amplitude, $J_1$ is a modified Bessel function, and $R_c$ is a radius of a pillbox cavity. $R_c$ is determined from the resonant frequency, i.e. $R_c = J_0(1) \, c/2\pi v$. For example, the stored energy of 200 and 800 MHz pillbox cavity are 313 and 20 Joules/m.

4.2 $10^{13}$ muons will generate about 1900 electron-ion pairs/cm in a 200 atm $H_2$ gas filled RF cavity. The nominal RF peak acceleration field in the RF energy recovery cavity is 15 ~ 20 MV/m. Thus, the *E/p* is 1.3 V/cm/Torr. From Figure 11, the expected *dw* in *E/p* = 1.3 is 2 $10^{-20}$ Joules/RF cycle. Therefore, the total RF power dissipation is dw × 1900 nμ × 100 cm = 0.16 Joules. On the other hand, a pure 200 atm $H_2$ gas filled RF cavity, if there is no recombination rate although it is not true, dw is 2 $10^{-17}$ Joules/cm/RF cycle. Then the RF power dissipation is 160 Joules. Figure 12 shows the estimate RF amplitude drop with 12 bunched muon beam with real muon collider beam parameter. The RF amplitude drops 5 and 8 % in 200 and 800 MHz cavities, respectively.

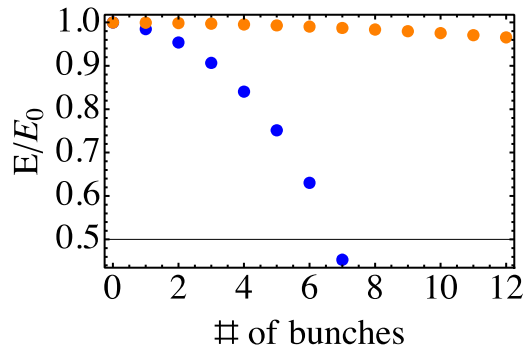

**FIGURE 7.** Estimated RF amplitude drop in a pure $H_2$ gas (blue) and an electronegative doped $H_2$ gas (orange) at gas pressure 200 atm.

## ACKNOWLEDGMENTS

We thank to Vladimir Shiltsev and Mark Palmer for supporting this program. We also great thank to the Fermilab Accelerator Division, the safety group, and the beam operator for helping this experiment.

This work is supported by FRA under DOE contract DE-AC02-07CH11359.